\newfont{\gothic}{eufb10 scaled\magstep 1}
\def\zid{\hbox{{1}\kern-.22em\hbox{l}}}
\documentstyle[12pt]{article}
\catcode`\@=11
\long\def\@makefntext#1{ 
\protect\noindent \hbox to 3.2pt {\hskip-.9pt
$^{{\ninerm\@thefnmark}}$\hfil}#1\hfill} 

\def\thefootnote{\fnsymbol{footnote}}
 \def\@makefnmark{\hbox to 0pt{$^{\@thefnmark}$\hss}}  

\def\ps@myheadings{\let\@mkboth\@gobbletwo
\def\@oddhead{\hbox{} 
\rightmark\hfil\ninerm\thepage}
\def\@oddfoot{}\def\@evenhead{\ninerm\thepage\hfil 
\leftmark\hbox{}}\def\@evenfoot{}
\def\sectionmark##1{}\def\subsectionmark##1{}}

\begin{document}

\newcommand{\symbolfootnote}{\renewcommand{\thefootnote}
	{\fnsymbol{footnote}}}
\renewcommand{\thefootnote}{\fnsymbol{footnote}}
\newcommand{\alphfootnote}
	{\setcounter{footnote}{0}
	 \renewcommand{\thefootnote}{\sevenrm\alph{footnote}}}
\newcounter{sectionc}\newcounter{subsectionc}\newcounter{subsubsectionc}
\renewcommand{\section}[1] {\vspace{0.6cm}\addtocounter{sectionc}{1}
\setcounter{subsectionc}{0}\setcounter{subsubsectionc}{0}\noindent
	{\bf\thesectionc. #1}\par\vspace{0.4cm}}
\renewcommand{\subsection}[1] {\vspace{0.6cm}\addtocounter{subsectionc}{1}
	\setcounter{subsubsectionc}{0}\noindent
	{\it\thesectionc.\thesubsectionc. #1}\par\vspace{0.4cm}}
\renewcommand{\subsubsection}[1]
{\vspace{0.6cm}\addtocounter{subsubsectionc}{1}
	\noindent {\rm\thesectionc.\thesubsectionc.\thesubsubsectionc.
	#1}\par\vspace{0.4cm}}
\newcommand{\nonumsection}[1] {\vspace{0.6cm}\noindent{\bf #1}
	\par\vspace{0.4cm}}
\newcounter{appendixc}
\newcounter{subappendixc}[appendixc]
\newcounter{subsubappendixc}[subappendixc]
\renewcommand{\thesubappendixc}{\Alph{appendixc}.\arabic{subappendixc}}
\renewcommand{\thesubsubappendixc}
	{\Alph{appendixc}.\arabic{subappendixc}.\arabic{subsubappendixc}}

\renewcommand{\appendix}[1] {\vspace{0.6cm}
        \refstepcounter{appendixc}
        \setcounter{figure}{0}
        \setcounter{table}{0}
        \setcounter{equation}{0}
        \renewcommand{\thefigure}{\Alph{appendixc}.\arabic{figure}}
        \renewcommand{\thetable}{\Alph{appendixc}.\arabic{table}}
        \renewcommand{\theappendixc}{\Alph{appendixc}}
        \renewcommand{\theequation}{\Alph{appendixc}.\arabic{equation}}
        \noindent{\bf Appendix \theappendixc #1}\par\vspace{0.4cm}}
\newcommand{\subappendix}[1] {\vspace{0.6cm}
        \refstepcounter{subappendixc}
        \noindent{\bf Appendix \thesubappendixc. #1}\par\vspace{0.4cm}}
\newcommand{\subsubappendix}[1] {\vspace{0.6cm}
        \refstepcounter{subsubappendixc}
        \noindent{\it Appendix \thesubsubappendixc. #1}
	\par\vspace{0.4cm}}

\def\abstracts#1{{
	\centering{\begin{minipage}{30pc}\tenrm\baselineskip=12pt\noindent
	\centerline{\tenrm ABSTRACT}\vspace{0.3cm}
	\parindent=0pt #1
	\end{minipage} }\par}}

\newcommand{\bibit}{\it}
\newcommand{\bibbf}{\bf}
\renewenvironment{thebibliography}[1]
	{\begin{list}{\arabic{enumi}.}
	{\usecounter{enumi}\setlength{\parsep}{0pt}
\setlength{\leftmargin 1.25cm}{\rightmargin 0pt}
	 \setlength{\itemsep}{0pt} \settowidth
	{\labelwidth}{#1.}\sloppy}}{\end{list}}

\topsep=0in\parsep=0in\itemsep=0in
\parindent=1.5pc

\newcounter{itemlistc}
\newcounter{romanlistc}
\newcounter{alphlistc}
\newcounter{arabiclistc}
\newenvironment{itemlist}
    	{\setcounter{itemlistc}{0}
	 \begin{list}{$\bullet$}
	{\usecounter{itemlistc}
	 \setlength{\parsep}{0pt}
	 \setlength{\itemsep}{0pt}}}{\end{list}}

\newenvironment{romanlist}
	{\setcounter{romanlistc}{0}
	 \begin{list}{$($\roman{romanlistc}$)$}
	{\usecounter{romanlistc}
	 \setlength{\parsep}{0pt}
	 \setlength{\itemsep}{0pt}}}{\end{list}}

\newenvironment{alphlist}
	{\setcounter{alphlistc}{0}
	 \begin{list}{$($\alph{alphlistc}$)$}
	{\usecounter{alphlistc}
	 \setlength{\parsep}{0pt}
	 \setlength{\itemsep}{0pt}}}{\end{list}}

\newenvironment{arabiclist}
	{\setcounter{arabiclistc}{0}
	 \begin{list}{\arabic{arabiclistc}}
	{\usecounter{arabiclistc}
	 \setlength{\parsep}{0pt}
	 \setlength{\itemsep}{0pt}}}{\end{list}}

\newcommand{\fcaption}[1]{
        \refstepcounter{figure}
        \setbox\@tempboxa = \hbox{\tenrm Fig.~\thefigure. #1}
        \ifdim \wd\@tempboxa > 6in
           {\begin{center}
        \parbox{6in}{\tenrm\baselineskip=12pt Fig.~\thefigure. #1 }
            \end{center}}
        \else
             {\begin{center}
             {\tenrm Fig.~\thefigure. #1}
              \end{center}}
        \fi}

\newcommand{\tcaption}[1]{
        \refstepcounter{table}
        \setbox\@tempboxa = \hbox{\tenrm Table~\thetable. #1}
        \ifdim \wd\@tempboxa > 6in
           {\begin{center}
        \parbox{6in}{\tenrm\baselineskip=12pt Table~\thetable. #1 }
            \end{center}}
        \else
             {\begin{center}
             {\tenrm Table~\thetable. #1}
              \end{center}}
        \fi}

\def\@citex[#1]#2{\if@filesw\immediate\write\@auxout
	{\string\citation{#2}}\fi
\def\@citea{}\@cite{\@for\@citeb:=#2\do
	{\@citea\def\@citea{,}\@ifundefined
	{b@\@citeb}{{\bf ?}\@warning
	{Citation `\@citeb' on page \thepage \space undefined}}
	{\csname b@\@citeb\endcsname}}}{#1}}

\newif\if@cghi
\def\cite{\@cghitrue\@ifnextchar [{\@tempswatrue
	\@citex}{\@tempswafalse\@citex[]}}
\def\citelow{\@cghifalse\@ifnextchar [{\@tempswatrue
	\@citex}{\@tempswafalse\@citex[]}}
\def\@cite#1#2{{$\null^{#1}$\if@tempswa\typeout
	{IJCGA warning: optional citation argument
	ignored: `#2'} \fi}}
\newcommand{\citeup}{\cite}

\def\fnm#1{$^{\mbox{\scriptsize #1}}$}
\def\fnt#1#2{\footnotetext{\kern-.3em
	{$^{\mbox{\sevenrm #1}}$}{#2}}}

\font\twelvebf=cmbx10 scaled\magstep 1
\font\twelverm=cmr10 scaled\magstep 1
\font\twelveit=cmti10 scaled\magstep 1
\font\elevenbfit=cmbxti10 scaled\magstephalf
\font\elevenbf=cmbx10 scaled\magstephalf
\font\elevenrm=cmr10 scaled\magstephalf
\font\elevenit=cmti10 scaled\magstephalf
\font\bfit=cmbxti10
\font\tenbf=cmbx10
\font\tenrm=cmr10
\font\tenit=cmti10
\font\ninebf=cmbx9
\font\ninerm=cmr9
\font\nineit=cmti9
\font\eightbf=cmbx8
\font\eightrm=cmr8
\font\eightit=cmti8
\hskip8cm ~ANL-HEP-CP-94-33%
\footnote{Talk by C. Zachos at PASCOS '94, Syracuse, NY, May 22, 1994.}
{}~~~~hep-th/9407044  \vskip12mm

\centerline{\tenbf THE PARADIGM OF PSEUDODUAL CHIRAL MODELS}
\vspace{0.8cm}
\centerline{\tenrm COSMAS K. ZACHOS }
\baselineskip=13pt
\centerline{\tenit High Energy Physics Division, Argonne National Laboratory,}
\baselineskip=12pt
\centerline{\tenit Argonne, IL 60439-4815, USA~~~(zachos@hep.anl.gov)}
\vspace{0.3cm}
\centerline{\tenrm and}
\vspace{0.3cm}
\centerline{\tenrm THOMAS L. CURTRIGHT}
\baselineskip=13pt
\centerline{\tenit Department of Physics, University of Miami, Box 248046}
\baselineskip=12pt
\centerline{\tenit  Coral Gables, Florida 33124,
USA~~~(curtright@phyvax.ir.Miami.edu)}
\vspace{0.9cm}
\abstracts{This is a synopsis and extension of Phys.~Rev.~{\em D49} 5408
(1994).
The {\em Pseudodual Chiral Model~}   illustrates 2-dimensional field theories
which possess an infinite number of conservation laws
{\em but also} allow particle production, at variance with naive
expectations---a folk theorem of integrable models.
We monitor the symmetries of the pseudodual model,
both local and nonlocal, as transmutations of the symmetries
of the (very different) usual {\em Chiral Model}.
We refine the conventional algorithm
to more efficiently produce the {\em nonlocal} symmetries of the model.
We further find the {\em canonical transformation} which connects the
usual chiral model to its
fully {\em equivalent dual model}, thus contradistinguishing the pseudodual
theory.}

\vfil
\twelverm
\baselineskip=14pt
\section{Introduction of the PCM and Outline of its Properties}
Many integrable models in two-dimensions evince the limiting feature of {\em
no particle production}, i.e.~complete elasticity.  There is a variant of the
$\sigma$-model for which this is not so, however (at least in perturbation
theory), the so-called {\em Pseudodual Chiral Model} of Zakharov and
Mikhailov\cite{Zak}, for which all interactions are distilled into a simple,
constant torsion term in the lagrangean; it amounts to a delicate
Wigner-In\"on\"u contraction of the target manifold in the WZW model in  which
the ``pion decay constant" is taken to infinity in tandem with the topological
integer coupling. The essential quantum features of the model were first
identified by Nappi\cite{nappi}, who calculated the nonvanishing $2\rightarrow
3$ production amplitude for this model, and who moreover demonstrated that the
model was inequivalent to the usual Chiral Model in its behavior under the
renormalization group: the Pseudodual Model is not asymptotically free. The
physics of the pseudodual model is very different from that of the usual chiral
model.

The models were previously compared within the framework of covariant path
integral quantization by Fridling and Jevicki, and similarly by Fradkin and
Tseytlin\cite{Fridling}. However, the focus of those earlier comparisons
was to exhibit (nonabelian-) dualized $\sigma$-models with torsion, which were
completely equivalent to the usual $\sigma$-model. Indeed, it was shown that a
model fully equivalent but dual to the usual Chiral Model could be constructed,
provided both nontrivial torsion and metric interactions were included in the
lagrangean.

Here, we focus on the differences between the Pseudodual Model and the usual
Chiral Model without enforcing equivalence. We investigate the Pseudodual Model
at the classical level and within the framework of canonical quantization, with
emphasis on the symmetry structure of the theory. We consider both local and
nonlocal symmetries, and compare with corresponding structures in the usual
Chiral Model. We present  a canonical transformation we have found, which
connects  the usual Chiral Model with its fully equivalent (nonabelian) dual
version, further clarifying the inequivalence of the pseudodual
theory\footnote{ An abelian penumbrance of this type of canonical
transformation has appeared recently  in the CERN preprint hepth/9406206 by
\'Alvarez, \'Alvarez-Gaum\'e, and Lozano.}. We provide a technically refined
algorithm for constructing the conserved nonlocal currents of the pseudodual
theory, an algorithm which is particularly well-suited to models with
topological currents for which the usual recursive algorithm temporarily stalls
at the lowest steps in the recursion before finally producing {\em genuine
nonlocals at the third step and beyond}. In the published paper, we also have
considered in detail the current algebra for the full set of local currents in
the pseudodual theory, which we omit here.  Other, related, more recent
investigations can be found in \cite{balog,fosco}.

The two-dimensional chiral model (CM) for matrix-valued fields $g$ is defined
by
$$
{\cal L}_1=\hbox{Tr}~\partial _\mu g\partial ^\mu g^{-1},
$$
with equations of motion which are conservation laws
$$
\partial _\mu J^\mu =0~~~~~\Longleftrightarrow ~~~~~\partial _\mu L^\mu =0.
$$
$J_\mu \equiv g^{-1}\partial _\mu g$ are the right-,
and $~L_\mu \equiv g\partial _\mu g^{-1}$ the left-rotation Noether currents
 of $G_{left}\times G_{right}$, respectively.
The pure-gauge form of these currents dictates that
 the non-abelian field-strength vanishes identically:
$$
{\partial }_\mu J_\nu -{\partial }_\nu J_\mu +[J_\mu ,J_\nu
]=0~~~~\Longleftrightarrow ~~~~\varepsilon ^{\mu \nu }{\partial }_\mu J_\nu
+\varepsilon ^{\mu \nu }J_\mu J_\nu =0~,
$$
and likewise for $L_\mu $. Such {\gothic curvature-free} local currents
underlie usual nonlocal-symmetry-generating
algorithms\cite{LuscherPohlmeyer,brezin,CurtrightZachos'80,Polyakov'80}.

The roles of current conservation and vanishing field
strength may be interchanged. A ``pseudodual''\cite{Fridling}
 transformation\cite{Zak,nappi}
leads to a {\em different model} for an antisymmetric matrix
field $\phi $. Define
$$
\label{curl}J_\mu =\varepsilon _{\mu \nu }\partial ^\nu \phi ,
$$
conserved identically. But now the
curvature-free condition above serves instead as the equation of
motion
$$
\label{motion}\partial ^\mu \partial _\mu \phi -{\textstyle \frac 12}%
\varepsilon _{\mu \nu }[\partial ^\mu \phi ,\partial ^\nu \phi ]=0,
{}~~~~~~~~~~~~~~~~~~~~~~~(\P)
$$
which follows from the lagrangean of the {\em Pseudodual Chiral Model} (PCM):
$$
{\cal L}_2=-{\textstyle \frac 14}\hbox{Tr}~\Bigl(\partial ^\mu \phi \partial
_\mu \phi +{\textstyle \frac 13}\phi \varepsilon _{\mu \nu }[\partial ^\mu
\phi ,\partial ^\nu \phi ]\Bigr).
$$

Nappi\cite{nappi} first observed that this model, in contrast to the Chiral
Model,  is anti-asymptotically free. Actually, this is now possible to
establish by inspection, given its subsumption in the general analysis of
$\sigma$-models with torsion\cite{BCZ}.
Introducing a (field-scale) coupling $\eta$ in the relative normalization
of the interaction
 term, one needs note the complete triviality of the metric (just the kinetic
term), $g_{ab} =\delta^{ab}$; the torsion $S_{abc} = \eta f_{abc} \sqrt{g}$
of the interaction term has now collapsed to a constant, merely the structure
constant times the coupling,
$S_{abc} = \eta f_{abc}= \eta \partial_{[a} e_{bc]}$,
for torsion potential $ e_{ab}= \eta f_{abc} \phi^c  $.
 This is, in fact, a limiting WZW model---a Wigner-In\"on\"u
contraction\cite{WI}
 of the group manifold such that the radius of the target hypersphere
(the ``pion decay constant") diverges {\em in tandem} with the integer WZW-term
coefficient. To one loop, Braaten, Curtright, and Zachos\cite{BCZ}
have shown that $e_{ab}$  evolves by the
antisymmetric part of the generalized Ricci tensor, vanishing in this case
of constant torsion, so   $e_{ab}$  does not renormalize.
In contrast,  $M {d \over dM} g_{ab} = -S_{acd} S_{b}^{~~cd} /2\pi =
-\eta^2 f_{acd} f_{b}^{~~cd} /2\pi =-\eta^2 \delta_{ab} C /2\pi$, where
$ C $ is the quadratic adjoint (dual-Coxeter/Casimir)  index, e.g.~$N-2$ for
O(N). Rescaling
the kinetic term to canonical normalization amounts to simply {\em increasing}
the interaction coupling as
$$
M {d \eta \over dM} ={3\eta \over 2} {C\eta^2  \over 2\pi}
={3\over 4 \pi} \eta^3 C,
$$
in agreement with the original direct calculation\cite{nappi}.

How do the fundamental symmetries generated by these and other
currents transmutate? Consider the conserved charge
$$
Q=\int \!dx\;J_0(x).
$$
For the CM, the time variation of $Q$ vanishes for
field configurations which extremize ${\cal L}_1$ by Noether's theorem;
while for the PCM, $Q=\phi (\infty )-\phi (-\infty) $, are time-independent for
{\em any} configurations with fixed boundary conditions ($\phi $ is
temporally constant at spatial infinity): $Q$ is a topological ``winding'' of
the field onto the spatial line and hence invariant under the continuous
flow of time.

The  $~~\phi \rightarrow O^T\phi O~~$  $G_{right}$-transformation
invariance of ${\cal L}_2$  yields the
(on-shell conserved) Noether currents
$$
R_\mu =[\phi ,\tilde J_\mu ]+{\textstyle \frac 13}[\phi ,[J_\mu ,\phi
]]=[\phi ,\partial _\mu \phi ]+{\textstyle \frac 13}\varepsilon _{\mu \nu
}[\phi ,[\partial ^\nu \phi ,\phi ]],
$$
where $\tilde J_\mu \equiv \varepsilon _{\mu \nu }\,J^\nu $. In contrast to
the CM, it is these currents, and not $J_\mu $, which generate
(adjoint) right-rotations in the PCM.

The PCM is also invariant under the
nonlinear symmetries\cite{nappi} $\phi\rightarrow \phi +\xi $ with Noether
currents
$$
Z_\mu =\tilde J_\mu +{\textstyle \frac 12}[J_\mu ,\phi ]=\partial _\mu \phi +%
{\textstyle \frac 12}\varepsilon _{\mu \nu }[\partial ^\nu \phi ,\phi ].
$$
The conservation law for these currents amounts to the
equations of motion Eq.($\P$) for the PCM (introduced as a
null-curvature condition for the topological $J_\mu $ currents of the model).
The equations of motion have been transmuted from conservation of $%
J_\mu $ for the CM to conservation of $Z_\mu$ for the PCM.
{These $Z_\mu $ currents are not curvature-free, however, but are
instead $J$-covariant-curl-free $\varepsilon ^{\mu \nu }\partial _\mu Z_\nu
+\varepsilon ^{\mu \nu }[J_\mu ,Z_\nu ]=0~. $ }
The currents $Z_\mu$ are contracted vestiges of the axial currents of the WZW
model, and we term them  ``pseudoabelian" since their charges commute among
themselves (more precicely, they close into the topological charge, vanishing
only for topologically trivial configurations), even
though this is not so for the entire current
algebra\cite{thing}. (Correspondingly, $J_\mu-R_\mu$ are vestiges of the vector
currents of the WZW model.)

These ``new'' local conserved
currents, $Z_\mu $ and $R_\mu $, are actually transmutations of the usual
first and second nonlocal currents of the CM, respectively. All
three sets of currents, $J_\mu ,Z_\mu ,R_\mu $, transform in the adjoint
representation of $O(N)_{right}$ (the charge of $R_\mu $).
The left-invariance $G_{left}$ has
degenerated: for the field $\phi $, left transformations are inert, and thus
right, or axial, or vector transformations are all indistinguishable.
The $G_{left}\times G_{right}$ symmetry of the chiral model,
the axial generators
of which are realized nonlinearly, has thus mutated in the PCM. On the one
hand it has been reduced by the loss of $G_{left}$, but on the other hand it
has been augmented by the nonlinearly realized pseudoabelian $Q_Z$ charges.

The left-currents $L_\mu$ of the CM don't generate left-rotations on the
PCM fields $\phi ,$ any more than the $J_\mu $
generate right-rotations. In the PCM, $L_\mu$ are  realized {\em nonlocally}:
{}~$\partial _\mu g=g~\varepsilon _{\mu \nu }\partial ^\nu \phi $, ~so~
$\partial _1g=g~\partial _0\phi ~$,
integrated at a fixed time,
$$
\label{g-as-fcn-of-phi}g(x,t)=g_0~P\exp (\int_x^\infty
dy~\partial _0\phi (y,t))~,
$$
assuming $g(\infty ,t)=g_0$. Consequently,
$$
L_\mu =g\partial _\mu g^{-1}=-g~(g^{-1}\partial _\mu g)~g^{-1}=-g~J_\mu
{}~g^{-1}
=-\varepsilon _{\mu \nu }~g~\partial ^\nu \phi ~g^{-1}=\qquad\qquad
$$
$$
{}~\qquad \qquad\qquad \qquad\qquad \qquad = \quad -\varepsilon _{\mu
\nu }\partial ^\nu (g~\phi ~g^{-1})+g~[\partial _\mu \phi ~,\phi ]~g^{-1}.
$$
These transform in the adjoint of $G_{left}$,
but these transformations only rotate the arbitrary
boundary conditions $g_0$, and do {\em not} affect $\phi $
at all. They thus commute with the right-rotations.
Discarding $g_0$ then banishes $G_{left}$ from the theory altogether.

None of the above results hinges  on the difference between
left- and right-currents. Left$\leftrightarrow $Right-reflected identical
results would have followed upon
interchange of left with right.

\section{Canonically Equivalent Dual $\sigma -$model}
The above nonlocal, invertible, fixed-time {\em map} relating
all $g$ and $\phi $ field configurations is, nevertheless,
{\gothic not a canonical  transformation}.
The quantum theories for ${\cal L}_1$ and ${\cal L}_2 $ are thus inequivalent
(e.g.~perturbation theory assumes canonical variables).
As an aside, we find instead a canonical transformation which maps the usual CM
onto an {\em equivalent} {\gothic  Dual Sigma Model} (DSM), with
torsion, different from the PCM, in broad agreement with the result
of conventional nonabelian duality transformations\cite{Fridling}.

E.g.~consider the standard O(4)~$\simeq $~O(3)$\times $O(3)
$ \simeq $   SU(2)$\times $SU(2) CM, with $g=\varphi ^0+i\tau ^j\varphi ^j$,
 ~~$\varphi ^0,\varphi^j~~(j=1,2,3)$, and $(\varphi ^0)^2+{\bf \varphi }^2=1$%
, where ${\bf \varphi }^2\equiv $ $\sum_{j\;}(\varphi ^j)^2$. Resolve
$\varphi ^0=\pm \sqrt{1-{\bf \varphi }^2}$, to get the CM,
$$
{\cal L}_1=\frac 12\left( \delta ^{ij}+\frac{\varphi ^i\varphi ^j}{1-{\bf %
\varphi }^2}\right) \partial _\mu \varphi ^i\partial ^\mu \varphi ^j.
$$
This is {\gothic canonically equivalent} to the DSM:
$$
\!\!\!\!
{\cal L}_3=\frac 1{1+4{\bf \psi }^2}\left( {\textstyle \frac 12}\left(
\delta ^{ij}+4\psi ^i\psi ^j\right) \partial _\mu \psi ^i\partial ^\mu
\psi ^j-\varepsilon
^{\mu \nu }\varepsilon ^{ijk}\psi ^i\partial _\mu \psi ^j\partial _\nu \psi
^k\right) ,
$$
which differs from the PCM, ${\cal L}_2$,
but reduces to it in the weak $\psi$ field limit,
i.e.~it contracts to it similarly
to the Wigner-In\"on\"u contraction of the WZW model. However, {\em no}
such canonical transformation may lead to the PCM instead.

The generator for a canonical
transformation relating $\varphi $ and $\psi $
at any fixed time is $F[\psi
,\varphi ]=\int_{-\infty }^\infty dx\;\psi ^iJ_i^1[\varphi ]$,
(where we choose\footnote{ N.B. Left-rotations on $\varphi $ alone do
nothing to this $F$;  $\psi ^i$ is a left-transformation singlet, just like its
conjugate quantity, $J_i^1[\varphi ]$, and $F[\psi ,\varphi ]$ is
left-invariant.}    ~the right, $V+A$, $J_\mu$),
$$
\label{F}F[\psi ,\varphi ]=\int_{-\infty }^{+\infty }dx~\psi ^i\;\left(
\sqrt{1-{\bf \varphi }^2}\frac {\stackrel{\leftrightarrow} {\partial}}
 {\partial x}\varphi ^i +\varepsilon
^{ijk}\varphi ^j\frac \partial {\partial x}\varphi ^k\right) .
$$

The conjugate momentum of $\psi ^i$:
\begin{eqnarray}
\label{Pi} \!\!\! \!\!\!
\pi _i=\frac{\delta F[\psi ,\varphi ]}{\delta \psi ^i}&=&
\sqrt{1-{\bf \varphi }^2}\;\frac \partial {\partial x}\varphi ^i
-\varphi ^i\frac \partial
{\partial x}\left( \sqrt{1-{\bf \varphi }^2}\right) +\varepsilon
^{ijk}\varphi ^j\frac \partial {\partial x}\varphi ^k = \nonumber \\
&=&\left( \sqrt{1-{\bf \varphi }^2}\;\delta ^{ij}+\frac{\varphi
^i\varphi ^j}{\sqrt{1-{\bf \varphi }^2}}-\varepsilon ^{ijk}\varphi ^k\right)
\frac \partial {\partial x}\varphi ^j=J_i^1 .\nonumber
\end{eqnarray}
The conjugate
of $\varphi ^i$:
\begin{eqnarray}
\label{PrelimVarPi}\!\!\!\!\!\!
\varpi _i=-\frac{\delta F[\psi ,\varphi ]}{\delta \varphi ^i}
&=&\left( \sqrt{1-
{\bf \varphi }^2}\;\delta ^{ij}+\frac{\varphi ^i\varphi ^j}{\sqrt{1-{\bf %
\varphi }^2}}+\varepsilon ^{ijk}\varphi ^k\right) \frac \partial {\partial
x}\psi ^j \nonumber \\
&+&\left( \frac 2{\sqrt{1-{\bf \varphi }^2}}\left( \varphi
^i\psi ^j-\psi ^i\varphi ^j\right) -2\varepsilon ^{ijk}\psi ^k\right) \frac
\partial {\partial x}\varphi ^j,\nonumber
\end{eqnarray}

Substitute for $\pi _i$ and $\varpi
_i$, in terms of $\frac \partial {\partial
t}\varphi ^j$ and $\frac \partial {\partial t}\psi ^j$, as follows from
${\cal L}_1$ and ${\cal L}_3$:
$$
\pi _i=\frac 1{1+4{\bf \psi }^2}\left( \left( \delta ^{ij}+4\psi ^i\psi
^j\right) \frac \partial {\partial t}\psi ^j+2\varepsilon ^{ijk}\psi ^j\frac
\partial {\partial x}\psi ^k\right) ,~~~~~~~~
\varpi _i=\left( \delta ^{ij}+%
\frac{\varphi ^i\varphi ^j}{1-{\bf \varphi }^2}\right) \frac \partial
{\partial t}\varphi ^j.
$$
The resulting covariant pair of first-order, nonlinear, partial differential
equations for $\varphi $ and $\psi $ constitute a B\"acklund transformation
connecting the two theories. Consistency of this B\"acklund transformation
is equivalent to the classical equations of motion for $\varphi $ and $\psi $.

Moreover, the relations
$$
\pi\cdot \pi  = \varphi'\cdot \varphi' + \frac {(\varphi\cdot \varphi')^2}
{1-\varphi^2}  ~,
$$
$$
\psi'\cdot \psi'= \varpi^2 -(\varphi\cdot \varpi)^2
+4 \pi^2 \psi ^2 -4(\pi\cdot \psi)^2- 4\sqrt{1-\varphi^2}~
\varepsilon^{ijk}\varpi_i \psi_j \pi_k
-4  \varphi\cdot \psi ~\varpi \cdot\pi
+4  \varphi\cdot \pi ~\varpi \cdot\psi ,
$$
$$
\varepsilon ^{ijk}\psi_i \pi_j \psi'_k  =
-2 \psi^2 \pi^2+2( \psi\cdot \pi)^2
+\sqrt{1-\varphi^2}
\varepsilon ^{ijk}\varpi_i \psi_j \pi_k
+\varphi\cdot  \psi \pi\cdot \varpi
-\varphi\cdot  \pi~ \psi\cdot \varpi ,
$$
may be combined to demonstrate the equivalence of the hamiltonian densities
in the respective theories:
$$
{\cal H}_3=
4\varepsilon ^{ijk}\psi_i \pi_j \psi_k'  +
\pi ^2 + \psi'\cdot\psi' +
4\psi^2 \pi^2
-4(\psi\cdot \pi)^2
=
\varpi ^2 - (\varphi\cdot \varpi)^2 + \varphi'\cdot\varphi' +
\frac {(\varphi\cdot \varphi')^2} {1-\varphi^2}
={\cal H}_1.
$$

Now, in the DSM, {\gothic what is the conserved, curvature-free current}? In
contrast
to the PCM, where it was essentially {\em forced} to be a topological
current, here a topological current by itself will not suffice; neither will
a conserved Noether current. (Under isospin transformations, $%
\delta \psi ^i=\varepsilon ^{ijk}\psi ^j\omega ^k$, the Noether
current of ${\cal L}_3$ is $I_i^\mu =\delta {\cal L}%
_3/\delta (\partial _\mu \omega ^i)$ so $I_i^0=\varepsilon ^{ijk}\psi
^j\pi _k$, but it is not curvature-free.)

Instead, the conserved,  curvature-free current
 ${\cal J}_i^\mu [\psi ,\pi ] =J_i^\mu [\varphi ,\varpi ]$
(identified with $J_i^\mu $ of  the CM) is a {\em mixture} of the Noether
isocurrent and a topological current:
${\cal J}_i^\mu =2I_i^\mu -\varepsilon ^{\mu \nu }\partial _\nu \psi ^i$, so
that ${\cal J}_i^1=\pi _i$. Both conservation and curvature-freedom now hold
on-shell.
$$
\label{PsiCurrent}{\cal J}_i^\mu =\frac{-1}{1+4{\bf \psi }^2}\left( \left(
\delta ^{ij}+4\psi ^i\psi ^j\right) \varepsilon ^{\mu \nu }\partial _\nu
\psi ^j+2\varepsilon ^{ijk}\psi ^j\partial ^\mu \psi ^k\right) .
$$
$$
\label{Jspace}{\cal J}_i^1\equiv \pi _i=\left( \sqrt{1-{\bf \varphi }^2}%
\;\delta ^{ij}+\frac{\varphi ^i\varphi ^j}{\sqrt{1-{\bf \varphi }^2}}%
-\varepsilon ^{ijk}\varphi ^k\right) \frac \partial {\partial x}\varphi
^j\equiv J_i^1,
$$
$$
\label{Jtime}{\cal J}_i^0\equiv -\frac \partial {\partial x}\psi
^i-2\varepsilon ^{ijk}\psi ^j\pi _k=-\sqrt{1-{\bf \varphi }^2}\;\varpi
_i-\varepsilon ^{ijk}\varphi ^j\varpi _k\equiv J_i^0.
$$

This last equation may also be integrated directly to yield
$\psi$ in terms of  $\varphi$, given the pure-gauge (zero curvature)
feature of    $J_\mu [\varphi]= g^{-1} \partial_{\mu} g $,
on which the canonical transformation was predicated:
$$
\frac  \partial {\partial x}\psi =\psi J_1 -J_1 \psi- J_0~~~~~~~\Longrightarrow
{}~~~~~\frac  \partial {\partial x}(g \psi g^{-1}) =
-g  J_0 g^{-1}  =  g \partial_0 g^{-1}~.
$$
The argument of the r.h.s.~has reduced to a {\em left} current component.
This equation readily  integrates to
$$
\psi  (x)=  g^{-1}(x) g(0)\psi  (0) g^{-1}(0)g(x)+
g^{-1}(x) \Bigl(\int_0^x dy g(y)    \partial_0 g^{-1}(y) \Bigr)  g(x)~~.
$$

N.B. Field-parity properties: under $\varphi \rightarrow -\varphi $,
the right current for the CM converts to
the left current, so that $F[-\psi,-\varphi]$ generates a canonical
transformation which projects onto right-invariants, instead.

\def\mapright#1{\vbox{\ialign{##\crcr
   $\hfil\scriptstyle{\ #1 \ }\hfil$
        \crcr\noalign{\kern+1pt\nointerlineskip}
     \rightarrowfill \crcr} }}
\def\mapdownl#1{\lower0.5ex\hbox{
      \llap{$\vcenter{\hbox{$\scriptstyle#1$}}$}}
           \lower0.5ex\hbox{\Big\downarrow}}
\def\mapdownr#1{\lower0.5ex\hbox{\Big\downarrow}
      \lower0.5ex\hbox{
         \rlap{$\vcenter{\hbox{$\scriptstyle#1$}}$}}}
\def\mapupl#1{\lower0.60ex\hbox{\llap{$\vcenter{\hbox{$\scriptstyle#1$}}$}}
           \lower0.5ex\hbox{\Big\uparrow}}
\def\mapupr#1{\lower0.5ex\hbox{\Big\uparrow}
         \lower0.60ex\hbox{\rlap{$\vcenter{\hbox{$\scriptstyle#1$}}$}}}
The connections among the four models discussed are summarized in the diagram:
$$
\matrix{
WZW & \mapright{contraction} & PCM~{\cal L}_2 \cr
\mapdownl{null~integer~coupling}& \  &  \mapupr{contraction} \cr
CM~{\cal L}_1&\leftarrow\!\!\!\mapright{\ canonical~equivalence \  } &DSM~{\cal
L}_3   ~~. \cr    }
$$
\phantom{.}

\section{Nonlocal Currents and Charges for the Pseudodual Model}
The full set of nonlocal conservation laws
\cite{LuscherPohlmeyer,brezin,CurtrightZachos'80,Polyakov'80,CurtrightZachos'93}
    follows from   any conserved, curvature-free currents such as $J_\mu $,
irrespective of the specific model considered. Introduce a dual
boost\cite{Zachos'80} spectral parameter $\kappa $ to define
$$
\label{C}C_\mu (x,\kappa )=-\frac{\kappa ^2}{1-\kappa ^2}J_\mu -\frac \kappa
{1-\kappa ^2}\;\tilde J_\mu ~~,
$$
where $\tilde J_\mu \equiv \varepsilon _{\mu \nu }\,J^\nu $.
Given these properties of $J_\mu$, it follows that
$$
\label{ZeroCurvC}\;\left( \partial ^\mu +C^\mu \right) \widetilde{C}_\mu
=0\;.
$$
This serves as the consistency condition for the two equations
$$
\label{PathOrdExp}\partial _\mu \,\chi ^{ab}(x)=-C_\mu ^{ac}\,\chi
^{cb}(x)\;,
$$
or, equivalently,
$$
\varepsilon _{\mu \nu }\partial ^\nu \chi =\kappa ~(\partial _\mu +J_\mu
)~\chi \;,
$$
which are solvable recursively\cite{brezin} in $\kappa $. Equivalently,
the solution $\chi $ can be expressed as a path-ordered
exponential (Polyakov's path-independent disorder\cite{Polyakov'80}
variable)
$$
\label{realPathOrdExp}\chi (x,\kappa )=P\exp \Bigl(-\int_{-\infty }^x
dy~C_1(y,t)\Bigr)\equiv 1\kern-0.36em\llap~1+\sum_{n=0}^\infty \kappa
^{n+1}\chi ^{(n)}\;.
$$

These ensure conservation of an {\gothic antisymmetrized} nonlocal ``master
current'':
$$
\label{NMC}\hbox{\gothic J}^\mu (x,\kappa )\equiv \frac 1{2\kappa
}~\varepsilon ^{\mu \nu }\partial _\nu \;\Bigl(\chi (x,\kappa )-\chi
^T(x,\kappa )\Bigr)\equiv \sum_{n=0}^\infty \kappa ^n~J_{(n)}^\mu (x)\;.
$$
The conserved master current acts as the generating functional of all
currents $J_{(n)}^\mu $ (separately) conserved order-by-order in $\kappa $.
E.g.~the lowest 4 orders yield:
\begin{eqnarray}
\kern-1.2em \hbox{\gothic J}_\mu(x,\kappa) &= & J_\mu(x)~
+~\kappa \Biggl( \tilde J_\mu(x)+{\textstyle \frac 12}
\Bigl[J_\mu(x)~,~\int_{-\infty}^x dy~ J_0(y)\Bigr] \Biggr)~+
\nonumber  \\
& &+~\kappa^2 \Biggl( \tilde J^{(1)}_\mu(x)+
{\textstyle \frac 12} (J_\mu(x)~\chi^{(1)} +
\chi^{(1)T}~J_\mu(x) ) \Biggr)~+ \nonumber  \\
& &+~\kappa^3 \Biggl( \tilde J^{(2)}_\mu(x)+
{\textstyle \frac 12}(J_\mu(x)~\chi^{(2)} +
\chi^{(2)T}~J_\mu(x)) \Biggr) \nonumber \\
& &+~{\cal O}(\kappa^4)~.\nonumber
\end{eqnarray}
This yields a conserved ``master charge''
$$
\label{WeDon'tTakeAmericanExpress}\hbox{\gothic G}(\kappa )=\int_{-\infty
}^{+\infty } dx~\hbox{\gothic J}_0(x,\kappa )\equiv
\sum_{n=0}^\infty \kappa ^n~Q_{(n)}\;.
$$
$Q_{(0)}$ is the conventional symmetry charge, while $Q_{(1)}$,
$Q_{(2)},Q_{(3)},...$ are the well-known nonlocal charges, best
studied for $\sigma -$models\cite{LuscherPohlmeyer,brezin,Polyakov'80}, the
Gross-Neveu model\cite{CurtrightZachos'81},
and supersymmetric combinations of the two\cite{CurtrightZachos'80}.

{\gothic However}, for the PCM,
$$
J_\mu ^{(0)}=J_\mu =\varepsilon _{\mu \nu }\partial ^\nu \phi
,~ ~~ ~~~\Longrightarrow ~ ~~~~~\chi ^{(0)}(x)=\phi (x)-\phi (-\infty ),
{}~~~~~\leadsto
$$
$$
\!\!\!\!\!\!
J_\mu ^{(1)}=\partial _\mu \phi +{\textstyle \frac 12}\varepsilon _{\mu \nu
}[\partial ^\nu \phi ,\phi ]-{\textstyle \frac 12}[J_\mu ,\phi (-\infty
)]=Z_\mu -{\textstyle \frac 12}[J_\mu ,\phi (-\infty )].
$$
Recall $\phi (-\infty )$ is taken to be
time-independent, and thus each piece of this current is separately
conserved. So, the CM$\leftrightarrow $PCM transmutation has yielded a
{\gothic local} current for the first nonlocal hopeful! Moreover,
$$
\kern-1.2em \chi ^{(1)}(x)=\int_{-\infty }^xdy~\Bigl(\partial _0\phi
(y)+\partial _1\phi (y)~\phi (y)\Bigr)-\phi (x)\phi (-\infty )+\phi (-\infty
)^2~.
$$
Likewise,  $~ J_\mu^{(2)}=$
$$
=\varepsilon_{\mu\nu} \partial^\nu \phi +[\partial_\mu \phi,\phi]
-\phi~ \varepsilon_{\mu\nu}\partial^\nu\phi~  \phi +
{\textstyle \frac 12}\varepsilon_{\mu\nu}\partial^\nu
\Bigl(\phi~ \chi^{(1)} + \chi^{(1)T} \phi \Bigr)
$$
$$
\qquad \qquad \qquad ~-{\textstyle \frac 12}[Z_\mu, \phi(-\infty)] +~
{\textstyle \frac 14}\varepsilon_{\mu\nu}\partial^\nu (
\phi^2 \phi(-\infty)+\phi(-\infty)\phi^2 ) ~~=
$$
$$
=
J_\mu -R_\mu -{\textstyle \frac 13} \varepsilon_{\mu\nu}\partial^\nu(\phi^3)~+
{\textstyle \frac 12}\varepsilon_{\mu\nu}\partial^\nu
\Bigl(\phi\chi^{(1)} + \chi^{(1)T} \phi \Bigr)
-{\textstyle \frac 12}[Z_\mu, \phi(-\infty)] \qquad
$$
$$
\qquad \qquad \qquad \qquad \qquad\qquad\qquad+~
{\textstyle \frac 14}\varepsilon_{\mu\nu}\partial^\nu (
\phi^2 \phi(-\infty)+\phi(-\infty)\phi^2 ) .
$$
On-shell properties of the currents have been used. However, this second
``nonlocal'' current is also effectively {\em local}: the skew-gradient term,
which might appear to contribute a nonlocal piece to the charge via $\chi
^{(1)}$, only contributes $[\phi (\infty ),Q_Z]/2$, i.e.~a trivial piece
based on a local current.

{\gothic But} the third step in the recursive algorithm is different:
$$
J_\mu ^{(3)}={\textstyle \frac 12}\Bigl(Z_\mu \chi ^{(1)}+\chi ^{(1)T}Z_\mu
\Bigr)+...~.
$$
(...) terms contribute only local pieces to the charge, whereas the term
written contributes ineluctable nonlocal pieces. Thus $J_\mu ^{(3)}$
is  {\em genuinely nonlocal}, like all higher currents. The action of
$Q^{(3)}$ (slightly improved to $Q_N$, as detailed below) on the field
changes the boundary
condition at $x=\infty $ to a different one than at $-\infty $, and thereby
switches its topological sector, which is quantified by $Q^{(0)}$:
$$
{}~ [\kern-0.45em\llap~[~ Q_N,\phi ^{ab}(y)~ ]\kern-0.45em\llap~]~ =-\bigl[%
[M^{ab},\phi(y)],\phi (y)\bigr]
+2\int_{-\infty}^{+\infty}dx~\varepsilon (y-x)[Z_0(x),M^{ab}], \;
$$
where ~~$(M^{ab})_{cd}\equiv \delta _{ac}\delta _{bd}-\delta _{ad}
\delta _{bc}$, ~and $
 [\kern-0.45em\llap~[ ~,]\kern-0.45em\llap~]$ represents Poisson brackets, in
contrast to matrix commutators $[,]$.

{\gothic In summary, for the pseudodual model, the charge $Q^{(0)}$ is
topological,
while $Q^{(1)}$ generates pseudoabelian shifts, $Q^{(2)}$
generates right rotations, and
$Q^{(n\geq 3)}$ appear genuinely nonlocal.}

\section{ Refinements and Remarks}
The above master current construction
starts off with a non-Noether (topological) current, then
``stalls'' twice at the first two steps before finally producing genuine
non-locals at the third step and beyond.
Here is an improved algorithm
which begins with the lowest nontopological (Noether) current $Z_\mu $ to
produce an alternate conserved master current which only stalls once.
Define\cite{CurtrightZachos'93}
$$
W_\mu (x,\kappa )\equiv Z_\mu +\kappa \tilde Z_\mu ,
$$
which is C-covariantly conserved:
$$
\partial ^\mu W_\mu +[C^\mu ,W_\mu ]=0~.
$$
This condition then empowers $W_\mu $ to serve as the seed for a new and
improved conserved master-current
$$
\hbox{\gothic W}_\mu (x,\kappa )=\chi ^{-1}W_\mu ~\chi
=Z_\mu +\kappa \left(T_\mu -[Z_\mu ,\phi (-\infty )]\right) +\qquad
\qquad \qquad
$$
$$
\qquad \qquad  +\kappa ^2\left( N_\mu -[T_\mu ,\phi
(-\infty )]+{\textstyle \frac 12}[[Z_\mu ,\phi (-\infty )],%
\phi (-\infty )]\right) +{\cal O}(\kappa ^3),
$$
where we have introduced
$$
\label{Tcurrent}T_\mu \equiv J_\mu -{\textstyle \frac 32}R_\mu =\tilde Z_\mu
+[Z_\mu ,\phi ],
$$
and where now the terms of second order and higher are genuinely nonlocal;
e.g.
$$
\label{Ncurrent}N_\mu =[T_\mu ,\phi ]-{\textstyle \frac 12}[[Z_\mu
,\phi ],\phi ]+[Z_\mu ,\int_{-\infty }^x\!\!dy~Z_0(y)]\;.
$$
This is a refined equivalent of $J_\mu ^{(3)}$ above. The terms in $%
\hbox{\gothic W}_\mu $ involving the constant matrices $\phi (-\infty )$ are
separately conserved.

In general, the seeds for such   improved
master currents only need be conserved currents, such as $Z_\mu $ above,
which also have a vanishing $J$-covariant-curl. E.g.~the previous
nonlocal currents themselves may easily be
fashioned to satisfy $J$-covariant-curl-free conditions and thereby seed
respective conserved master currents.

In summary, at tree level (and thus for massless excitations),
it has been made evident  that particle
production is not prevented by nonlocal conservation laws,
as holds for the CM\cite{LuscherPohlmeyer}, but is often thought to
automatically occur in general\cite{balog}. In our paper, we
further work out the current algebra of the currents discussed, and we moreover
list the known {\em local } sequence of conserved currents predicated on
conserved, curvature-free currents such as $J_\mu $. But, in this case,
elasticity  theorems\cite{Parke} on the prevention of particle production as a
consequence  of  Lorentz tensor charges such as those are evaded, since they
require massive states, which are absent at the semiclassical level considered
 here.

\section{Acknowledgements}
Work supported by the NSF grant
PHY-92-09978 and
the U.S.~Department of Energy, Division of High Energy Physics, Contract
W-31-109-ENG-38.

\end{document}